\documentclass[twocolumn,aps,showpacs,preprintnumbers,amsmath,amssymb]{revtex4}
\usepackage{graphicx}
\usepackage{bm}

\begin{document}

\title{Multiple-scale analysis and renormalization for pre-asymptotic 
scalar transport}

\author{A. Mazzino$^{1}$, S.~Musacchio$^{2}$ and A.~Vulpiani$^{2,3}$}

\affiliation{$^{1}$ INFM--Department of Physics, 
University of Genova and INFN sezione di Genova, Via Dodecanneso 33, 
I--16146 Genova, Italy.}
\affiliation{$^{2}$ Department of Physics and INFM (UdR and CSM), 
University La Sapienza, P.le A. Moro 2, I-00185 Roma, Italy.}
\affiliation{$^{3}$ INFN - Sezione di Roma La Sapienza.}

\date{\today}

\begin{abstract}
Pre-asymptotic transport of a scalar quantity passively
advected by a velocity field formed by
a large-scale component superimposed to
a small-scale fluctuation is investigated both analytically and by
means of numerical simulations.
Exploiting the multiple-scale expansion one arrives at a Fokker--Planck
equation which describes the pre-asymptotic scalar dynamics. Such
equation is associated to a Langevin equation involving a
multiplicative noise and  an effective (compressible) drift.
For the general case, no explicit expression for both the  
effective drift and the effective diffusivity (actually a tensorial
field) can be obtained. We discuss an approximation under which
an explicit expression for the diffusivity (and thus for the drift) 
can be obtained. Its expression permits to  highlight
the important fact that the diffusivity explicitly depends on the
large-scale advecting velocity.
Finally, the robustness of the aforementioned
approximation is checked numerically by means of direct numerical 
simulations.
\end{abstract}


\maketitle

\section{Introduction}
Many problems, from biology to geophysics, include a variety of
degrees of freedom with very different time scales~\cite{1}.
As important examples of systems with multiple time scales we can
mention the protein folding and the climate.
While the time scale for the vibration of the covalent bonds is
$O(10^{-15} s)$, the folding time for the proteins may be of the
order of seconds~\cite{2}.
In an analogous way, climate dynamics involves processes with
characteristic times ranging from days (atmosphere)
to $10^2-10^3$ years (deep ocean and ice shields)~\cite{3}.

Even modern supercomputers are not able to simulate all the relevant
scales involved in such difficult problems.
Consequently, scientists concerned with multiple time scale systems
must develop suitable techniques for the treatment of the ``slow
dynamics'' in terms of effective equations~\cite{1}.
This is a very old problem: an early example of such techniques
is the averaging method in mechanics.
Starting from a systems on $2N$ ordinary differential equations
written in the angle-action variables, where the angles
$(\theta_1$,$\theta_1$,\dots,$\theta_N)$ are ``fast'' and the actions
$(I_1$,$I_2$,\dots,$I_N)$ are ``slow'', the averaging method
gives the leading order behavior of the actions by an effective
equation for the averaged quantities 
$( \langle I_1 \rangle$,$\langle I_2 \rangle$,\dots,$\langle I_N \rangle )$
obtained averaging on the angles.

For the sake of self-consistency, we briefly recall the general problem.
Let us limit ourselves to  systems with sole two times scales and
denote by
$\bm{x}$ and $\bm{y}$ the slow degrees of freedom and 
the fast ones, respectively. The time evolution is given by a set
of ordinary differential equations:
\begin{equation}
\frac{d \bm{x}}{ d t}=\bm{f}(\bm{x}, \bm{y})
\label{1.1a}
\end{equation}
\begin{equation}
\frac{d \bm{y}}{d t} = \frac{1}{\epsilon} \bm{g}(\bm{x}, \bm{y})
\label{1.1b}
\end{equation}
where $\epsilon \ll 1$ is the ratio between the fast and slow
characteristic time scales.
The main goal is to approximate the motion of the slow variables
$\bm{x}$  by an effective equation where  the fast
variables $\bm{y}$ do not appear.

Up to now, different methods have been proposed. Among the many,
we can mention the Mori-Zwanzig formalism~\cite{4,4b}, invariant manifolds,
averaging methods~\cite{5}, conditional expectations~\cite{6} 
and Langevin equations~\cite{7,8}.

Following the seminal works on the Brownian motion~\cite{aa,aaa}, 
it seems rather natural to mimick the dynamics of fast
variables, $\bm{y}$, through
a white-in-time process, which amounts to describing the slow variables
$\bm{x}$ in terms of a suitable Langevin equation.
This approach is at the basis of the
seminal paper of Hasselmann in climate modeling in terms
of stochastic equations~\cite{7}.

Under rather general conditions~\cite{9}, one has the result 
that in the limit of small $\epsilon$ the slow dynamics is ruled by a 
Langevin equation with multiplicative noise:
\begin{equation}
\frac{d \bm{x}}{d t} = \bm{f}_{eff}(\bm{x})
+ \bm{\sigma}(\bm{x}) \bm{\eta} 
\label{1.2}
\end{equation}
where $\bm{\eta}$ is a white-noise vector, i.e.
the components are Gaussian processes such that
$ \langle \eta_i(t) \rangle=0$ ,
$ \langle \eta_i(t) \eta_j(t') \rangle = \delta_{ij}\delta(t-t')$
and $\bm{\sigma}(\bm{x})$ is a tensorial field.\\
This class of problems attracts a great deal of attention in many 
field of physics, including, e.g. statistical physics.
We just mention the celebrated renormalization group which can be seen
as a technique to explicitly determine $\bm{f}_{eff}(\bm{x})$
and $\bm{\sigma}(\bm{x})$ in Hamiltonian systems~\cite{10}.\\
There are rather general results~\cite{Kurtz,Papanicolaou}
which give explicit expression for the coefficients in Eq.~(\ref{1.2})
in terms of expectations over the fast process generated by Eq.~(\ref{1.1b}) 
at slow $\bm{x}$ fixed. On the other hand there are 
technical difficulties in the practical use of such results, 
and therefore approximations based on physical ideas 
(as those in \cite{4,4b,5,6,7}) are required; 
for a recent review see~\cite{GKS04}. 
Another interesting approach is to use the theoretical results 
in~\cite{Kurtz,Papanicolaou} to built and test a numerical strategy 
for the effective computation with Eq.~(\ref{1.2})
\cite{VE03,FVE04}.

The aim of this paper is the investigation of a specific class 
of problems with many active, coupled degrees of freedom. To be more
specific, we focus our attention on 
the large-scale transport
of a scalar field, $\theta(\bm{x},t)$, advected by an incompressible
velocity field consisting in the superposition of
a large scale, slowly varying, component and a small scale, rapidly varying,
fluctuation. Namely,
\begin{equation}
\partial_t\theta(\bm{x},t) + \bm{v}(\bm{x},t)\cdot\bm{\nabla} \theta(\bm{x},t)
=D_0 \Delta \theta(\bm{x},t) 
\label{FP}
\end{equation}
with
\begin{equation}
\bm{v}(\bm{x},t)=\bm{U}(\bm{x},t)
+\varepsilon \bm{u}(\bm{x},t)
\label{vare}
\end{equation}
where the typical length-scale of $\bm{U}$  and $\bm{u}$
are $L$ and $\ell$, respectively, and $\ell/L \ll1$.
The parameter $\varepsilon$ controls the relative strength
of velocity components.
It is worth recalling that 
Eq.~(\ref{FP}) is nothing but the Fokker--Planck equation
associated to the Langevin equation:
\begin{equation}
\frac{d \bm{x}(t)}{d t} =\bm{v}(\bm{x},t)+\sqrt{2 D_0} \eta \,\, .
\label{langevin}
\end{equation}

Our main aim here is to consider an
effective large-scale transport equation for
the large-scale scalar field, $\theta_L$, varying on scales of
order of $L$, in which
the dynamical effects of the smallest scales appear  
via a  renormalized (enhanced) diffusivity. 
Such an equation reads \cite{M97}
\begin{equation}
\partial_t \theta_L(\bm{x},t)
+\bm{U}(\bm{x},t)\cdot\bm{\partial} \theta_L(\bm{x},t)
=\partial_{i}
\left[ \partial_j D_{ij}(\bm{x},t) \theta_L(\bm{x},t) \right] 
\label{large_FP}
\end{equation}
or, in the equivalent form,
\begin{equation}
\partial_t \theta_L(\bm{x},t)
+ \bm{\partial}\cdot \left[ \bm{U}^E (\bm{x},t) \theta_L(\bm{x},t)\right ]
=\partial_i\partial_j \left [D^E_{ij}(\bm{x},t) \theta_L(\bm{x},t) \right] 
\label{large_FP2}
\end{equation}
where
\begin{equation}
U^E_i(\bm{x},t)\equiv \left [U_i(\bm{x},t) + \partial_j D_{i j}(\bm{x},t)
\right ]
\label{ueff}
\end{equation}
\begin{equation}
D^E_{ij}\equiv
\frac{D_{ij}+D_{ji}}{2}\; .
\label{dsimm}
\end{equation}
We anticipate that $D_{ij}(\bm{x},t)$
is in general neither symmetric nor positive defined. 
Its symmetric part (which is also positive defined) contributes
to the diffusion process while both the symmetric and the antisymmetric
parts enter, in general, in the effective advection velocity $\bm{U}^E$ which
turns out to be compressible. As we will show, we have identified a
sufficient condition  which rules out the antisymmetric 
contribution of $D_{ij}(\bm{x},t)$ in $\bm{U}^E (\bm{x},t)$. In this
case, $D_{ij}^E(\bm{x},t)$ is the only relevant
(in general unknown) field of the problem.\\
The Eulerian view for the large-scale dynamics given by
(\ref{large_FP2}) is equivalent to the Lagrangian description
(written in the Ito formalism):
\begin{equation}
\frac{d \bm{x}(t)}{d t}=\bm{U}^E(\bm{ x},t)+
\sqrt{2 D^{E}_{ij}(\bm{ x},t) } \eta  \;.
\end{equation}

Unfortunately, although we know the equation for the pre-asymptotic
dynamics of a scalar field, no explicit expression for 
$D_{ij}(\bm{x},t)$ is available in general. We will discuss in the paper
how to proceed perturbatively in $\varepsilon$ [the parameter 
defined in (\ref{vare})]
in order to obtain an approximate explicit expression for
$D_{ij}(\bm{x},t)$. Other than for applicative purposes, the advantage
of this expression is that it permits to highlight the important
result that  $D_{ij}(\bm{x},t)$ explicitly depends on the large-scale
advection $\bm{U}$. This is unlike the common way to think an
eddy-diffusivity contribution as the result of interactions involving
the sole small scales.\\
Finally, we will show the results of direct numerical simulations (DNS)
in order to assess the robustness of the approximation and thus
of the underlying physical mechanisms at the basis of the
dependence of $D_{ij}(\bm{x},t)$ on the large-scale velocity field.

In more detail, the paper is organized as follow.
In Sec.~\ref{mse} we will show how to derive~(\ref{large_FP})  exploiting the
multiple-scale strategy (see, e.g., \cite{BLP78,Piretal,M97,BCVV95,11}).
The latter is a renormalized perturbation method which requires to
have $\ell/L \ll1$. In general, the determination of the effective
parameters can be performed only numerically (see, e.g., \cite{M97,11}).
If, in addition to $\ell/L \ll1$,
we also assume  $\varepsilon\ll 1$, an explicit expression for 
$D_{ij}(\bm{x},t)$ can be derived.
Some important conclusions can be drawn.
A part the (trivial) case
of shear flow,  $D_{ij}(\bm{x},t)$
cannot be constant;
the components of the diffusivity tensor depend 
on the large-scale velocity as well as on 
the small scales. This latter point seems to be relevant
for geophysical applications where such dependence on large-scale
flow is often not considered. 

In  Sec.~\ref{pertu} we will compute 
$D_{ij}(\bm{x},t)$ perturbatively in $\varepsilon$.
Only the leading, $O(\varepsilon)$,  term of the series 
will be calculated  analytically. Such term is exact in some
particular cases. Although also for the 
higher order terms analytical expressions can be given, their
 complexity do not permit to extract relevant informations.\\
Numerical simulations performed on the exact Eq.~(\ref{FP})
show that the approximate first-order solution
is in very good agreement with numerical simulations also
for $\varepsilon$ and $\ell/L$ not too small, say $0.2 - 0.4$.
In addition we propose an empirical ``recipe'' 
to obtain a constant eddy-diffusivity for pre-asymptotic transport.
This is shown in Sec.~\ref{num}.

In Sec.~\ref{rg} we will discuss  how, at least in principle, in the
presence of velocity fields $\bm{u}(\bm{x},t)$ containing
contributions at many different scales, the multiple-scale
approach can be iterated, in a way that a renormalization
group  procedure naturally emerges with the result that  an effective
equation for asymptotic scales which involves
an effective diffusivity, $D^{E}$,
can be obtained. Because of technical difficulties, the explicit detailed 
computation of the iteration procedure appear quite cumbersome. Nevertheless,
for the dependence of $D^{E}$ on the velocity field, one can derive 
(and generalize) some results previously obtained in a phenomenological way.

Finally, Sec.~\ref{conclu}  is reserved for final conclusions and open problems.

\section{Multiple-scale analysis}
\label{mse}
Multiple-scale analysis  applied to transport phenomena 
(see e.g. Ref.~\cite{BLP78})
constitutes a powerful tool to extract the equations ruling the 
large-scale dynamics from first principles, i.e., the equations
describing the entire set of spatial/temporal degrees of freedom.\\
From a general point of view, the large-scale equations involve
renormalized parameters which can usually be determined by solving
an auxiliary differential problem which requires the knowledge 
of the fully resolved fields. This is for instance the case
analyzed in Ref.~\cite{Piretal} where it is shown that
the large-scale dynamics of a scalar field, in the
presence of scale separation 
with respect to the (small scale) advecting velocity field, 
is governed by an effective equation which is always diffusive.
The diffusion coefficient (actually a tensor) turns out to be 
larger than the bare (molecular) diffusion coefficient:
unsolved turbulent motion enhances the large-scale transport
(see Ref.~\cite{BCVV95}). \\
The latter result has been generalized in Ref.~\cite{M97} where
the pre-asymptotic passive scalar dynamics has been analyzed.
There, the assumption of dealing with a small-scale advecting velocity
field has been relaxed and the possible dependence of velocity on
scales comparable with those of the scalar has been taken into
account. As a results, we will show here that
the large-scale (pre-asymptotic) equation does not have a 
Fokker--Planck structure although it involves 
a renormalized diffusivity (actually a tensorial field).
The latter is varying on scales
comparable with those of the large-scale components of the advecting 
velocity. As a consequence, no Lagrangian description is 
associated to such Eulerian equation.\\ 

\subsection{Pre-asymptotic dynamics of a passive scalar: heuristic 
considerations}
\label{prea}
The starting point of our analysis is the equation ruling the
evolution of a passive scalar field, $\theta({\bm x},t)$,
in an incompressible velocity field $\bm{v}$:
\begin{equation}
\partial_t\theta(\bm{x},t)+\bm{v}\cdot \bm{\nabla}
\theta(\bm{x},t)=
D_0\Delta\theta (\bm{x},t) \;\;\; .
\label{FPbis}
\end{equation}
If one is interested to study the scalar dynamics in the deep
infra-red limit (i.e. very large scales) the proper choice for $\bm{v}$
is as in Ref.~\cite{BCVV95}: a small-scale field varying on scales
well separated from those at which the scalar dynamics is observed.\\
More frequently, in real applications (e.g., in geophysics) 
one could be interested to study the scalar dynamics on large scales
where the advecting velocity is however still relevant (i.e., on such
wave-numbers the velocity is appreciably non-zero). Following Ref.~\cite{M97}, 
the simplest  way to treat a similar situation is to decompose $\bm{v}$
as the sum of  $\bm{u}(\bm{x},t)$ and $\bm{U}(\bm{x},t)$. The former
is assumed to vary on what we call ``small scales'' [i.e. wave-numbers
of $O(1)$]
while the latter evolves on ``large scales'' having wave-numbers 
of $O(\epsilon)$, the same 
at which we aim at investigating the scalar dynamics. \\
Naive arguments would suggest a simple (wrong) conclusion: 
$\bm{U}(\bm{x},t)$ gives the advection contribution in the large-scale
equation for $\theta$ while the renormalized diffusion coefficient
emerges from small-scale interactions between $\theta$ and $\bm{u}$.
A detailed analysis actually shows that such conclusion is wrong:
the large-scale velocity, $\bm{U}(\bm{x},t)$, is not responsible
for the sole large scale advection, but it also enters in the 
renormalized diffusivity.\\
Before proceeding with a formal derivation where this effect clearly
emerges, let us give an heuristic argument in favor of such a
mechanism.\\
Suppose to have a large-scale initial condition for $\theta$  at time $t=0$
behaving on wave-numbers of $O(\epsilon)$
and, moreover,  $\bm{v}=\bm{u}$ 
(i.e. the case discussed in Ref.~\cite{BCVV95}). 
Due to the advection term
$\bm{u}\cdot \bm{\nabla}$ in (\ref{FP}), scalar components with
wave-numbers  of $O(1+\epsilon)$
are excited at larger times. The latter scalar components can interact, 
again due to the action
of $\bm{u}\cdot \bm{\nabla}$, with those of $\bm{u}$ 
to generate, at successive times,
large-scale components of $\theta$.  This is the basic mechanisms
giving raise to the renormalization of the bare diffusion
coefficient via interaction involving small scales. \\
Let us now repeat the argument in the presence of $\bm{U}$ which
varies on wave-numbers of $O(\epsilon)$. The interactions we have
described above continue to work with the main difference that
new contributions to the wave-numbers
of $O(1+\epsilon)$ now come from interactions of 
$O(\epsilon)$ modes of $\bm{U}$ and $O(1)$ modes of $\theta$.
The latter modes being involved in the renormalization
process, one can conclude that $\bm{U}$ plays a role in such
renormalization.
Whether or not this is really the case requires  a formal analysis,
which is the subject of next section.

\subsection{Formal analysis for the pre-asymptotic scalar transport}
\label{formal}
Following Ref.~\cite{M97}, let us decompose $\bm{v}$ as
$\bm{v}(\bm{x},t)\equiv \bm{U}(\bm{x},t) + \bm{u}(\bm{x},t)$
where $\bm{U}(\bm{x},t)$ and $\bm{u}(\bm{x},t)$
are assumed to be periodic in boxes
of sides $O(\epsilon^{-1})$ and $O(1)$, respectively.
(The technique we are going to describe can be extended 
with some modifications to  handle the case of a random, 
homogeneous and stationary velocity field).\\
Our focus is on the large-scale  dynamics 
of the field $\theta (\bm{x},t)$ on spatial scales of 
$O(\epsilon^{-1})$.\\
In the spirit of multiple-scale analysis, 
we introduce a set of {\em slow variables}
${\bm X}=\epsilon {\bm x}$, $T={\epsilon}^{2} t$ and
$\tau=\epsilon t$
in addition to the {\em fast variables} $({\bm x},t)$. 
The scaling of the times $T$ and $\tau$  
are suggested by physical reasons: we are searching for diffusive 
behavior on large time scales of $O(\epsilon^{-2})$ 
taking into account the effects played by the advection contribution
occurring on time scales of $O(\epsilon^{-1})$.\\
The prescription of the technique is to treat the variables
as independent. It then follows that
\begin{equation}
\label{derivate}
\partial_i\mapsto\partial_i+\epsilon\nabla_i\,;\qquad
\partial_t\mapsto\partial_t+\epsilon\partial_{\tau} +\epsilon^2\partial_T,
\end{equation}
\begin{equation}
\label{campi}
\bm{u}\mapsto\bm{u}(\bm{x},t)\,;\qquad \bm{U}\mapsto\bm{U}(\bm{X},T)
\end{equation}
where $\partial$ and $\nabla$ denote the derivatives with respect to
fast and slow space variables, respectively. The solution is sought as a
perturbative series
\begin{equation}
\label{serie}
\theta({\bm x},t;{\bm X},T;\tau)=\theta^{(0)}+\epsilon\theta^{(1)}+
\epsilon^2\theta^{(2)}+\ldots ,
\end{equation}
where the functions $\theta^{(n)}$ depend, {\it a priori}, on both fast
and slow variables. By inserting (\ref{serie}) and (\ref{derivate})
into (\ref{FP}) and equating terms having equal powers in $\epsilon$, 
we obtain a hierarchy of equations in which both fast
and slow variables appear. The solutions of interest to us are those
having the same periodicities as the velocity field, $\bm{u}(\bm{x},t)$.

By averaging such equations 
over the small-scale periodicity (here denoted by
$\langle\cdot\rangle$), 
a set of equations involving the sole
large-scale fields (i.e. depending on ${\bm X}$,  $T$ and
$\tau$) are easily obtained. Obviously, 
such equations must be solved recursively,
because of the fact that solutions of a given order appear as coefficients
in the equations at the higher orders. Let us show in detail this 
point.\\

It is not difficult to verify that the equations at order $\epsilon$ and 
$\epsilon^2$ read \cite{M97}:
\begin{align}
O(\epsilon):\ \ \ &
\begin{array}{ll} 
\displaystyle{\partial_t\theta^{(1)} + (\bm{ v} \cdot
\bm{\partial})\,\theta^{(1)}- D_0 \,\partial^2\theta^{(1)}=} \\
\displaystyle{-(\bm{v}\cdot\bm{\nabla})\theta^{(0)}
-\partial_{\tau}\theta^{(0)}}  
\label{o1} 
\end{array}
\\
%
O(\epsilon^2):\ \ \ &
\begin{array}{ll}
\displaystyle{\partial_t\theta^{(2)} + (\bm{ v} \cdot
\bm{\partial})\,\theta^{(2)}- D_0 \,\partial^2\theta^{(2)}=} \\
\displaystyle{-\partial_T\theta^{(0)}-(\bm{v}\cdot\bm{\nabla})\theta^{(1)}+
D_0\nabla^2\theta^{(0)} } \\
\displaystyle{+2D_0(\bm{\partial}\cdot\bm{\nabla})\theta^{(1)}-
\partial_{\tau}\theta^{(1)}} \;\;\; .
\label{o2}
\end{array}
\end{align}
The linearity of  (\ref{o2}) permits to search for a solution
in the form
\begin{eqnarray}
\theta^{(1)}(\bm{ x},t;\bm{ X},T;\tau)&=&\langle\theta^{(1)}\rangle
(\bm{ X},T;\tau)\nonumber\\
&+&\bm{ \chi}(\bm{x},t;\bm{X},T)\cdot\bm{\nabla}
\theta^{(0)}(\bm{ X},T;\tau)  \;\;\; ,
\label{sol2}
\end{eqnarray}
where $\theta^{(0)}$ depends on the sole large-scale variables
as in Ref.~\cite{BCVV95}. 
Plugging (\ref{sol2})
into the solvability condition for (\ref{o2}),
one obtains  the equation
\begin{equation}
\label{r2_2}
\partial_T\theta^{(0)}+ (\bm{U}\cdot\bm{\nabla})\langle 
\theta^{(1)}\rangle
+\partial_{\tau}\langle\theta^{(1)}\rangle =
\nabla_i \left (D_{ij}  
\nabla_j \theta^{(0)}\right ) 
\end{equation}
where 
\begin{equation}
D_{ij}(\bm{X},T)=\delta_{ij}D_0- \langle u_{i} 
\chi_{j} \rangle
\label{meglio} 
\end{equation}
is a second-order tensorial field
and $\bm{\chi}(\bm{ x},t ;\bm{ X},T)$ 
has a vanishing average over the periodicities and 
satisfies the following equation:
\begin{equation}
\label{chi}
\partial_t { \chi_j}+\left[(\bm{ u}+ \bm{ U})\cdot\bm{ \partial}\right]
{ \chi}_j - D_0\, \partial^2{ \chi}_j = -{u}_j \;\;\; .
\end{equation}

Note that,
when $\bm{U}$ is not a pure mean flow but depends on 
$\bm{X}$ and $T$, the equation (\ref{chi}) must  be solved
for each value of $\bm{X}$ (and eventually $T$).

From Eq.~(\ref{r2_2}) 
and from the solvability condition of Eq.~(\ref{o1}),
\begin{equation}
\partial_{\tau}\langle\theta^{(0)}\rangle
+(\bm{U}\cdot\bm{\nabla})\, \langle\theta^{(0)}\rangle = 0 ,
\label{r1}
\end{equation}
one obtains the equation for the large-scale field $\theta_L$ defined as: $\theta_L\equiv 
\langle\theta^{(0)}\rangle+\epsilon \langle \theta^{(1)}\rangle$:
\begin{equation}
\partial_t\theta_{L} + (\bm{U} \cdot \bm{\partial})\,
\theta_{L}=
\partial_i \left (D_{ij}  \partial_j
\theta_{L}\right )
    \label{lsse}
\end{equation}
where the usual variables $\bm{x}, t$ are used.

The important point to note is that $D_{ij}$ is in general neither
symmetric nor positive defined.  On the contrary, 
it is easy to show \cite{M97},
that $D^E_{ij}\equiv (D_{ij}+D_{ji})/2 $ is (obviously) symmetric 
and positive defined. Its expression can immediately be obtained 
from (\ref{chi}) in term of the sole auxiliary field :
\begin{equation}
D^E_{ij}=D_0\langle \partial_p\chi_i \partial_p\chi_j \rangle \; .
\end{equation}
In terms of $D^E_{ij}$ and 
$D^A_{ij}\equiv (D_{ij} - D_{ji})/2 $, the pre-asymptotic equation
(\ref{lsse}) takes the form
\begin{equation}
\partial_t\theta_{L} + \bm{\partial}\cdot (\bm{U}^E\theta_L )
=
\partial_i\partial_j \left (D^E_{ij} \theta_{L}\right ).
    \label{lsse-bis}
\end{equation}
where
\begin{equation}
U^E_i(\bm{x},t)\equiv \left [U_i(\bm{x},t) + \partial_j D_{i
j}^E(\bm{x},t)+ \partial_j D_{i j}^A(\bm{x},t)
\right ]
\end{equation}
is an effective compressible advecting velocity \cite{PK02}. 
Advection by compressible velocities have been 
investigated, e.g., in Refs.~\cite{M03} and \cite{VA97}.

\subsection{Formal analysis for the asymptotic scalar transport}
\label{formal_as}

Our aim is now to investigate transport on 
scales much larger than the typical length of the field
${\bm U}$, i.e. on scales ${\cal L} \gg L$.\\
Homogenization leads to a purely diffusive 
dynamics which involves a set of new  slow variables 
${\bm {\mathcal{X}}} = \epsilon' {\bm X}$ and ${\mathcal{T}} = 
\epsilon^{'\,2} T$ describing 
the large-scale  field 
$\theta_{\mathcal{L}} = \langle \theta_L \rangle$. Averages are now 
over the cell of  size $L$:
\begin{equation}
\partial_{\mathcal{T}} \theta_{\mathcal{L}} = 
D^{\mathcal{L}}_{ij} \nabla_i \nabla_j \theta_{\mathcal{L}} \;.
\label{eq:3.11}
\end{equation}
There are two different ways to arrive at the large-scale 
equation (\ref{eq:3.11}). The first way is to apply the 
homogenization technique from Eq.~(\ref{lsse}) while the second
possibility is to start directly from Eq.~(\ref{FP}). Let us 
consider the first option. In this case,
the asymptotic  eddy-diffusivity tensor $D^{\mathcal{L}}$ will then result
from the combined effects of the advection given by the large-scale 
flow $\bm{U}({\bm X},T)$ and the diffusion at scale $\ell$ which 
also depend on space and time through $D_{ij}({\bm X},T)$,
\begin{eqnarray} 
D^{\mathcal{L}}_{ij} &=& 
- \frac{\langle U_i \chi_j \rangle + \langle U_j \chi_i \rangle}{2} 
+ \frac{\langle D_{ik} \partial_k \chi_j \rangle +
   \langle D_{jk} \partial_k \chi_i \rangle}{2}\nonumber\\
&&+ \frac{\langle D_{ij} \rangle + \langle D_{ji} \rangle}{2} ,
\label{eq:3.12}
\end{eqnarray}
where the vector field ${\bm \chi}$ is here solution
of the auxiliary equation
\begin{equation}
\partial_t \chi_k + ({\bm U} \cdot {\bm \partial} ) \chi_k 
- \partial (D_{ij} \partial_j \chi_k) = - U_k + \partial_i D_{ik} \;.
\label{eq:3.13}
\end{equation}

If one follows the second way to obtain the 
large-scale equation (\ref{eq:3.11}),
the (exact) value of the eddy-diffusivity tensor, $D^{{\mathcal L},ex}$, 
depends on both the  
molecular diffusivity and the advection by the total velocity field
$ {\bm v} = {\bm U} + {\bm u}$: 
\begin{equation}
D^{{\mathcal L},ex}_{ij} = \delta_{ij} D_0 
- \frac{\langle v_i \chi_j \rangle + \langle v_j \chi_i \rangle}{2} . 
\label{eq:3.14}
\end{equation}
Here, the auxiliary field ${\bm \chi}$ 
is the solution of the following equation
\begin{equation}
\partial_t {\bm \chi} + ({\bm v} \cdot {\bm \partial} ){\bm \chi} -
D_0 \partial^2 {\bm \chi} = - {\bm v} \;.
\label{eq:3.15}
\end{equation}
The latter procedure gives the exact value of the 
eddy-diffusivity tensor $D^{{\mathcal L},ex}$, 
but requires the detailed knowledge of the velocity field at 
both large and small scales. On the other hand, the expression 
obtained from Eq.~(\ref{eq:3.12}) (which, in general, does not
coincide with $D^{{\mathcal L},ex}$) 
 is based on the sole large-scale 
velocity, $\bm{U}$, and the effects of the small-scale flow are included 
in the eddy-diffusivity $D_{ij}({\bm X},T)$. \\
A clear indication that $D^{{\mathcal L},ex}\neq D^{\mathcal L}$ can
be obtained by noting that 
the eddy-diffusivity tensor, $D_{ij}$,
does not depend on the relative position (i.e. possible spatial shifts)
between the two fields ${\bm U}$ and ${\bm u}$. This is an obvious
consequence of scale separation which washes out all detailed
differences between the two fields. Therefore, 
the effects of relative shifts between ${\bm U}$ and ${\bm u}$ 
which are taken into account in the exact
eddy-diffusivity tensor  $D^{{\mathcal L},ex}$ are missed 
by the approximate expression for
$D^{\mathcal{L}}$.
A comparison between the expressions for the asymptotic diffusivities
obtained by following  the two different homogenization procedures
allows one  to quantify the error of the approximate strategy.

It is worth mentioning a particular case in which 
both procedures leads to the same results. This is the case
when the velocity field ${\bm v}$ is
given by the sum of two parallel steady shears,
\begin{equation}
{\bm v} ({\bm x};{\bm X}) = {\bm u} ({\bm x}) + {\bm U} ({\bm X}) ,
\label{eq:3.16}
\end{equation}
with 
\begin{equation}
{\bm u} (\bm{x}) = (u(y),0) \;, 
{\bm U} (\bm{X}) = (U(Y),0) 
\label{eq:3.17}
\end{equation}
where $U$ and $u$ vary on scales of order of $L$ and $\ell$, respectively.

A first homogenization on the small scales $\ell$ 
leads to an eddy-diffusivity (we use the equivalent notations
$D_{xx}\equiv D_{11}$ and $D_{yy}\equiv D_{11}$)
\begin{equation}
D_{xx} = D_0 + 
\frac{1}{2} \int \frac{|\hat{u}|^2 dk}{D_0 k^2} \;, 
D_{yy} = D_0 \;, D_{i j} = 0 \; \forall i \ne j \;. 
\label{eq:3.18}
\end{equation}
One can now repeat the same homogenization
procedure at large scales $L$, obtaining 
\begin{eqnarray}
D^{\mathcal L}_{xx} &=& D_{xx} + 
\frac{1}{2} \int \frac{|\hat{U}|^2 dk}{D_{yy} k^2} \nonumber \\
&=& D_0 + 
\frac{1}{2} \int \frac{|\hat{u}|^2 dk}{D_0 k^2} +
\frac{1}{2} \int \frac{|\hat{U}|^2 dk}{D_0 k^2} \;,
\label{eq:3.19}
\end{eqnarray} 
which coincides with the exact 
coefficient obtained from the homogenization
carried out from Eq.~(\ref{FP}) which involves
the total velocity field ${\bm v} = {\bm U} +{\bm u}$:
\begin{equation}
D^{{\mathcal L},ex}_{xx} = D_0 + 
\frac{1}{2} \int \frac{(|\hat{U}|^2 + |\hat{u}|^2) dk}{D_0 k^2} \;. 
\label{eq:3.20}
\end{equation}

\section{An approximate expression for the eddy-diffusivity field}
\label{pertu}
In previous section we have shown how to reduce the
computation of the eddy-diffusivity tensor $D_{ij}(x,t)$ 
to the solution of an auxiliary equation.
It is however worth noting that the 
parametric dependence on the large scale variables ${\bm X},T$
in the auxiliary field ${\bm \chi}({\bm x},t; { \bm X},T)$ in Eq.~(\ref{chi})
imposes a rather severe limit to the 
practical use of Eq.~(\ref{lsse}). If the large scale velocity ${\bm U}$ 
depends on space and time,  one has indeed to solve an auxiliary equation 
in $(2d+1)$ dimensions. 

Therefore, except very few cases in which one can obtain an analytic solution
for ${\bm \chi}({\bm x},t; { \bm X},T)$,
e.g. in the case of orthogonal shears (see Sec.~\ref{orto}), 
Eq.~(\ref{meglio}) does not provide a practical tool for evaluating 
the eddy-diffusivity of generic flows. The computational cost 
required for the solution of the auxiliary equation can indeed be heavier  
than that required for the solution of the complete equation.  

In the following we will show how the presence of an intense large-scale flow 
permits to overcome this limit. 
Indeed, if the strength of the large scale flow ${\bm U}$ 
is much larger than that of the small scale velocity field ${\bm u}$, 
one can seek the solution 
of the auxiliary equation as a  perturbative series in 
the small parameter $\varepsilon = u/U$:
\begin{equation}
{\bm \chi}({\bm x},t;{\bm X},T) = 
{\bm \chi}^{(0)} + \varepsilon {\bm \chi}^{(1)} + 
\varepsilon^2 {\bm \chi}^{(2)} +\dots \;, 
\label{eq:3.1}
\end{equation}
where the functions ${\bm \chi}^{(n)}$ depend on both fast and slow variables.
By inserting Eq.~(\ref{eq:3.1}) into Eq.~(\ref{chi}) and equating 
terms having equal powers in ${\varepsilon}$, we obtain a 
hierarchy of equations:
\begin{eqnarray}
\partial_t {\bm \chi}^{(0)} + 
({\bm U} \cdot {\bm \partial}) 
{\bm \chi}^{(0)}
- D_0 \partial^2 {\bm \chi}^{(0)}  & = & 0 \;, \label{eq:3.2} \\
\partial_t {\bm \chi}^{(1)} + 
({\bm U} \cdot {\bm \partial}) 
{\bm \chi}^{(1)}
- D_0 \partial^2 {\bm \chi}^{(1)}  & = &  -{\bm u} \;, \label{eq:3.3} \\
\cdots \cdots \phantom{-----------}&   & \nonumber \\
\partial_t {\bm \chi}^{(n)} + 
({\bm U} \cdot {\bm \partial}) 
{\bm \chi}^{(n)}
- D_0 \partial^2 {\bm \chi}^{(n)}  & = &  -({\bm u} \cdot {\bm \partial})  
                                          {\bm \chi}^{(n-1)} \;. \nonumber \\ 
\phantom{-----------}              &   &  \label{eq:3.4}
\end{eqnarray}
The $0$-th order equation has the trivial solution 
${\bm \chi}^{(0)} = {\bm \chi}^{(0)}({\bm X},T)$, 
which clearly  
does not contribute [see Eq.~(\ref{meglio})] to $D_{ij}({\bm X},T)$, 
while the higher order equations can be easily solved in Fourier space.
At first order in $\varepsilon$ the solution reads
\begin{equation}
\hat{\bm \chi}^{(1)}({\bm k},\omega;{\bm X},T) = 
\frac{- \hat{\bm u}({\bm k},\omega)} 
{i(\omega + {\bm U} \cdot {\bm k}) + k^2 D_0}
\label{eq:3.5}
\end{equation}
which, exploiting (\ref{meglio}), leads to the following expression 
\begin{eqnarray}
D_{ij}({\bm X},T) &=& D_0 \delta_{ij} \nonumber\\
                  &+& \int d{\bm q} \; d\omega 
                      \left\{ \frac{\mbox{Re} \left[ 
                      \hat{u}_{i}(-{\bm q},-\omega) 
                      \hat{u}_{j}( {\bm q}, \omega) \right] 
                      q^2 D_0}
                      {(\omega + {\bm U} \cdot {\bm q})^2 + q^4 D_0^2} 
                      \right. \nonumber\\
                  &+& \left.  \frac{\mbox{Im} \left[ 
                      \hat{u}_{i}(-{\bm q},-\omega) 
                      \hat{u}_{j}( {\bm q}, \omega) \right] 
                      (\omega+{\bm U} \cdot {\bm q})}
                      {(\omega + {\bm U} \cdot {\bm q})^2 + q^4 D_0^2} 
                      \right\} \nonumber \\
                  &+& O (\varepsilon^3) \;.
\label{eq:3.6}
\end{eqnarray}

Eq.~(\ref{eq:3.6}) permits to highlight some important points.
The eddy-diffusivity is not simply determined by the small-scale flow: 
it actually has an explicit dependence on the large-scale velocity components. 
A rough estimation of the eddy-diffusivity based  
on the sole small-scale field can lead to completely wrong results 
when a large-scale flow is present.
Moreover, the variation in space and time of the velocity field 
${\bm U}({\bm X},T)$ 
induces an implicit dependence on 
the slow variables ${\bm X},T$ in the eddy-diffusivity, 
which thus becomes a tensorial field.  
We stress the fact that such a dependence on ${\bm X},T$
is not a consequence of the 
approximation~(\ref{eq:3.6}), the same properties holds 
if one use the exact ${\bm \chi}$.  \\
The physical origin of this effect is the strong sweeping 
caused by the large-scale velocity field, which  
changes the effective correlation times of the small-scale flow. 
Therefore, the frequencies $\omega$ which appear in Eq.~(\ref{eq:3.5}) 
experiences a Doppler-shift corresponding to
the inverse of the sweeping time ${\bm U} \cdot {\bm k}$. 
Only when the temporal variation of the small-scale flow is much 
faster than the large-scale sweeping, 
i.e. when the power spectrum of the small-scale flow is peaked at very
high frequencies,  $\omega \gg {\bm U} \cdot {\bm k}$,
one obtains a constant tensor
which does not depend on ${\bm U}$: 
\begin{eqnarray}
D_{ij}({\bm X},T) &=& D_0 \delta_{ij} \nonumber\\
                  &+& \int 
                      \left\{ \frac{\mbox{Re} \left[ 
                      \hat{u}_{i}(-{\bm q},-\omega) 
                      \hat{u}_{j}( {\bm q}, \omega) \right] 
                      q^2 D_0}
                      {\omega^2 + q^4 D_0^2} 
                      \right. \nonumber\\
                  &+& \left.  \frac{\mbox{Im} \left[ 
                      \hat{u}_{i}(-{\bm q},-\omega) 
                      \hat{u}_{j}( {\bm q}, \omega) \right] 
                      \omega}
                      {\omega^2 + q^4 D_0^2} 
                      \right\} d{\bm q} \; d\omega \nonumber \\
                  &+& O (\varepsilon^3) \;.
\label{eq:3.10}
\end{eqnarray}

As we have already shown  in Sec.~\ref{formal}, both the symmetric 
and the antisymmetric part of $D_{i j}$ contribute to the effective
advecting velocity
\begin{equation}
U^E_i(\bm{x},t)\equiv U_i(\bm{x},t) + \partial_j D_{i
j}^E(\bm{x},t)+ \partial_j D_{i j}^A(\bm{x},t) \; .
\end{equation}
Exploiting the explicit expression for $D_{i j}$ it is easy to derive 
a sufficient condition under which $D_{i j}^A$ is identically zero
[and thus  $\partial_j D_{i j}^A(\bm{x},t)=0$]. If such condition is
satisfied, then the sole $D_{i j}^E$ is relevant for the dynamics at
pre-asymptotic scales. This seems interesting for applications
in view of the fact that, in three dimensions, only six, rather than
nine, fields (the components of $D_{i j}^E$)  must be taken into account.

Formally, the analytic result obtained for the eddy-diffusivity 
is valid only in the limits $\ell / L \ll 1$ and $u / U \ll 1 $.
Therefore we must expect some discrepancies between the actual results
for $\ell / L \sim 1$ and $u / U \sim 1 $ 
and those obtained exploiting the multiple-scale method.
Actually, we will show in Sec.~\ref{num}
that good results are obtained even if the
ratio between the characteristic 
scales of the flows, $\epsilon = \ell / L$, and amplitudes, 
$\varepsilon = u/U $,  are not too small. 

\subsection{A sufficient condition for the effective advecting velocity}
As a starting point let us
rewrite $\hat{u}_{i}$ in term of its real and imaginary parts,
$\hat{u}_{i}$ = $\hat{u}_{i}^{\tiny\mbox{R}} +\sqrt{-1}
\hat{u}_{i}^{\tiny\mbox{I}}$, and plug it into
(\ref{eq:3.6}) which takes the form:
\begin{eqnarray}
D_{ij}({\bm X},T) &=& D_0 \delta_{ij} \nonumber\\
                  &+& \int \left\{ \frac{q^2 D_0}
                      {(\omega + {\bm U} \cdot {\bm q})^2 + q^4 D_0^2}
                      \right. \nonumber\\
             &\times& \left[
                      \hat{u}_{i}^{\tiny\mbox{R}}({\bm q},\omega)
                      \hat{u}_{j}^{\tiny\mbox{R}}({\bm q},\omega)
                      +
                      \hat{u}_{i}^{\tiny\mbox{I}}({\bm q},\omega)
                      \hat{u}_{j}^{\tiny\mbox{I}}({\bm q},\omega)
                      \right]  \nonumber\\
                  &+& \left[ 
                      \hat{u}_{i}^{\tiny\mbox{R}}({\bm q},\omega)
                      \hat{u}_{j}^{\tiny\mbox{I}}({\bm q},\omega)
                      - 
                      \hat{u}_{i}^{\tiny\mbox{I}}({\bm q},\omega)
                      \hat{u}_{j}^{\tiny\mbox{R}}({\bm q},\omega)
                      \right] \nonumber\\
             &\times& \left. \frac{\omega+{\bm U} \cdot {\bm q}}
                      {(\omega + {\bm U} \cdot {\bm q})^2 + q^4 D_0^2} 
                      \right\} d{\bm q} \; d\omega \nonumber \\
                  &+& O (\varepsilon^3) \;.
\label{eq:3.6bis}
\end{eqnarray}

From the above expression one immediately realizes that  the
antisymmetric part of $D_{ij}({\bm X},T)$ is
\begin{eqnarray}
D_{ij}^A({\bm X},T) &=& \int                                
\left[ \hat{u}_{i}^{\tiny\mbox{R}}({\bm q}, \omega)
       \hat{u}_{j}^{\tiny\mbox{I}}({\bm q}, \omega) -
       \hat{u}_{i}^{\tiny\mbox{I}}({\bm q}, \omega)
       \hat{u}_{j}^{\tiny\mbox{R}}({\bm q}, \omega) \right] 
       \nonumber \\ 
                   &\times& \frac{\omega + {\bm U} \cdot {\bm q}}
                   {(\omega + {\bm U} \cdot {\bm q})^2 + q^4 D_0^2} 
                   \; d{\bm q} \; d\omega \;.
\end{eqnarray}
The condition for the latter to be zero is
\begin{equation}
       \hat{u}_{i}^{\tiny\mbox{R}}({\bm q}, \omega)
       \hat{u}_{j}^{\tiny\mbox{I}}({\bm q}, \omega) -
       \hat{u}_{i}^{\tiny\mbox{I}}({\bm q}, \omega)
       \hat{u}_{j}^{\tiny\mbox{R}}({\bm q}, \omega) =0 \; ,
\label{suff}
\end{equation}
from which sufficient conditions for its validity are immediately obtained:
\begin{equation}
\hat{\bm u}^{\tiny\mbox{R}}({\bm q}, \omega)=0\quad 
\forall{\bm q}~\mbox{and}~\omega \quad \mbox{or}\quad
\hat{\bm u}^{\tiny\mbox{I}}({\bm q}, \omega)=0 \quad
\forall{\bm q}~\mbox{and}~\omega\; .
\label{2suff}
\end{equation}
Conditions (\ref{2suff}) amount to saying that if the 
small-scale velocities
have defined parity with respect to space/time inversion, then
the sole symmetric part
of $D_{i j}$ controls the pre-asymptotic scalar dynamics.

To conclude, it is worth observing that the formula (\ref{eq:3.6})
can be generalized to random small-scale velocity mimicking 
turbulent small-scale fluctuations. In this case 
$\hat{u}_{i}(-{\bm q}, -\omega) \hat{u}_{j}({\bm q}, \omega)$
in (\ref{eq:3.6}) must be replaced by 
$\langle \hat{u}_{i}(-{\bm q}, -\omega) \hat{u}_{j}({\bm q}, \omega)
\rangle$ 
where brackets denote the average 
with respect to small-scale velocity statistics.  If one deals
with stationary, homogeneous and isotropic fluctuations the spectral
tensor $\langle \hat{u}_{i}(-{\bm q}, -\omega) \hat{u}_{j}({\bm q},
\omega) \rangle$ is invariant under ${\bm q}\to - {\bm q}$ and
$\omega\to -\omega$ with the immediate consequence that
\begin{equation}
\langle \hat{u}_{i}^{\tiny\mbox{R}}({\bm q}, \omega)
       \hat{u}_{j}^{\tiny\mbox{I}}({\bm q}, \omega)\rangle -
     \langle  \hat{u}_{i}^{\tiny\mbox{I}}({\bm q}, \omega)
       \hat{u}_{j}^{\tiny\mbox{R}}({\bm q}, \omega)\rangle =0 \; ,
\label{suffturbo}
\end{equation}
a condition which generalizes (\ref{suff}).

\subsection{The case of orthogonal shears}
\label{orto}
Although Eq.~(\ref{eq:3.6}) is just a first order approximation, it
provides a concrete tool to estimate the eddy-diffusivity, 
and it can be shown that for the particular case of 
orthogonal shears it recovers the exact solution~\cite{M97}. 
Indeed, if the velocity field is the sum of two orthogonal shears
\begin{equation}
{\bm v} ({\bm x},t;{\bm X},T) = {\bm u} ({\bm x},t) + {\bm U} ({\bm X},T)
\label{eq:3.7}
\end{equation}
with 
\begin{equation}
{\bm u} ({\bm x},t) = (u(y,z,t),0,0) \;, 
{\bm U} ({\bm X},T) = (0,U(X,Z,T),0) 
\label{eq:3.8}
\end{equation}
it follows from (\ref{chi}) that
the unique non-vanishing component of the auxiliary field is the one 
in the direction of the small-scale velocity, and it 
is constant along that direction. 
Therefore the small-scale velocity field does not 
give contributions in the advective term of Eq.~(\ref{chi}) which can exactly 
be  solved in Fourier space to obtain
\begin{equation}
\hat{\bm \chi}({\bm k},\omega;{\bm X},T) = 
\left(
\frac{- \hat{\bm u}({\bm k},\omega)}{
i(\omega + {\bm U} \cdot {\bm k}) + k^2 D_0} 
,0,0 \right) \; .
\label{eq:3.9} 
\end{equation}

\section{Numerical results and ``empirical recipes'' for the pre-asymptotic transport}
\label{num}
In the previous section we have discussed 
a perturbative solution and its possible limitations when 
$u/U$ and $\ell / L$ are not very small. 
Let us now present some numerical results and an empirical ``recipe''
for a constant (i.e without space and time dependence) pre-asymptotic eddy-diffusivity. 

\subsection{Numerical results}

As an example of small-scale incompressible flows 
we consider a steady cellular flow~\cite{sg88,11,BCVV95}
defined by the stream function
$\psi = \psi_0 \sin(kx) \sin(ky)$ with $\psi_0 = u/k$:
\begin{eqnarray}
{\bm u} & = & (\partial_y \psi, -\partial_x \psi) 
\nonumber \\
& = &
(u \sin(kx) \cos(ky), - u \cos(kx) \sin(ky)) \;.
\label{eq:3.23}
\end{eqnarray}
Its characteristic length-scale is given by $\ell = 2 \pi / k$ 
and its amplitude is $u$. 

In the absence of large-scale velocity fields and for 
large Peclet numbers ($Pe = u \ell / D_0$), 
it is possible to show by means of simple physical arguments~\cite{p85} 
that this periodic array of small vortexes 
give rise to an enhancement of the 
effective diffusivity $D^{E} \sim D_0 \sqrt{Pe}$.  
A precise estimation of this constant eddy-diffusivity can be obtained 
by the numerical solution of Eq.~(\ref{chi}), with ${\bm U}=0$.

\begin{figure}
\begin{center}
\includegraphics[scale=0.7]{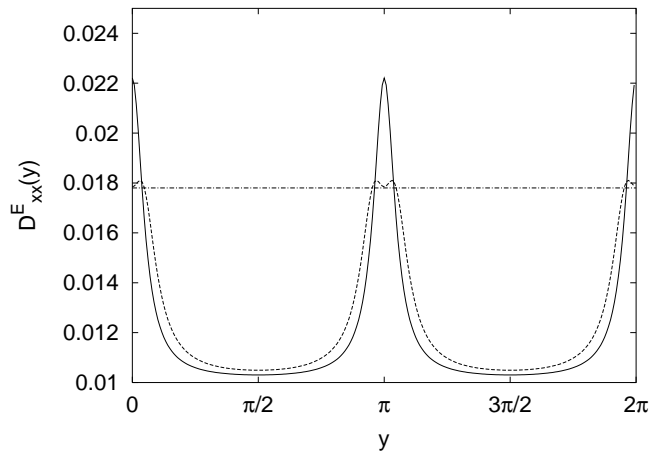}
\caption{The eddy-diffusivity $D^E_{xx}(y)$ 
resulting from a small-scale cellular flow superimposed to a large-scale 
shear in the $x$ direction. The approximation~(\ref{eq:3.28}) 
(solid line) recover quite well the exact multiple-scale solution 
(dashed line), 
except for narrow regions where the large-scale flow vanishes ($y=n \pi$) 
and the actual diffusivity recovers the constant estimation 
based on the sole small-scale cellular flow (dash-dotted line).
The parameter values are $U=1$, $L=2 \pi$, $u/U=1/4$, $\ell/L =1/8$,
$D_0=0.01$. Units are made dimensionless according to Eq.~(\ref{eq:adim}).
} 
\label{fig:1}
\end{center}
\end{figure}

The modifications induced on the eddy-diffusivity by the presence of 
a large-scale flow, 
\begin{equation}
{\bm U} = (U(X,Y,T),V(X,Y,T)) \; ,
\label{eq:3.24}
\end{equation}
with characteristic length-scale $L = \ell / \epsilon$ and strength 
$U = u / \varepsilon$ can be estimated from Eq.~(\ref{eq:3.6}). 
Thanks to the simplicity of our small-scale flow, the integral in 
Eq.~(\ref{eq:3.6}) reduces to the sum of contributions 
of four modes $(\pm k, \pm k)$,
and trivial calculations lead to 
\begin{eqnarray}
D^{E}_{ij} &=& D_0 \delta_{ij} \left\{ 1+ \frac{1}{4} u^2 \left[
\frac{1}{(U + V)^2 + (2 k D_0)^2 } 
\right. \right. \nonumber \\
&&
\left. \left.
+ 
\frac{1}{(U - V)^2 + (2 k D_0)^2 } \right] \right\}
+ O(\varepsilon^3)
\label{eq:3.25}
\end{eqnarray}
For such a system the antisymmetric part $D^{A}_{ij}$ is identically zero.

\begin{figure}[pt]
\begin{center}
\includegraphics[scale=0.7]{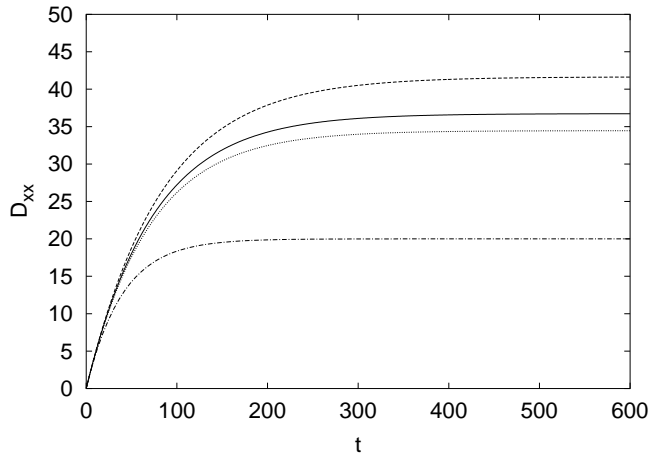}
\caption{Time evolution of the asymptotic eddy-diffusivity 
in the direction of the large-scale shear.
up to its convergence to its constant value.
The scale separation between the large-scale shear and the 
small-scale cellular flows is $\ell / L  = 1/4$, 
the ratio of amplitudes is $u / U = 1/4$ and 
the molecular diffusivity is fixed to the value $D_0 = 10^{-2}$.
The first-order approximation in $\varepsilon = u/U$ (solid line) 
provides a good estimation on the actual values, which depend on the 
relative phase-shift between the two fields: case $a$ 
is denoted by dashed line, case $b$ is denoted by dotted line. 
For comparison we also show the results 
obtained from the ``naive estimation'' (dash-dotted line) in which the 
effects on the large-scale flow have been neglected.
Units are made dimensionless according to Eq.~(\ref{eq:adim})
}
\label{fig:2}
\end{center}
\end{figure}

Let us now focus on two idealized large-scale flows, 
which are representative of two broad classes 
of realistic situations: 
a steady shear 
\begin{equation}
{\bm U} = (U \sin(Ky),0)  
\label{eq:3.26}
\end{equation}
and a large-scale replica of the cellular flow 
\begin{equation}
{\bm U} = (U \sin(Kx) \cos(Ky), -  U \cos(Kx) \sin(Ky) )  \; .
\label{eq:3.27}
\end{equation}
Their characteristic length-scale is $L = 2 \pi / K$ 
and $U$ is their amplitude.
For the case of the large-scale shear, Eq.~(\ref{eq:3.25})
reduces to 
\begin{equation}
D^{E}_{ij} = D_0 \delta_{ij} \left( 1+ \frac{1}{2}
\frac{u^2}{U^2 \sin^2(y) + (2 k D_0)^2 } \right)
+ O(\varepsilon^3)
\label{eq:3.28}
\end{equation}
while in the case of the large-scale cellular flow one gets
\begin{eqnarray}
D^{E}_{ij} & = &  D_0 \delta_{ij} \left\{ 1+ \frac{1}{4} u^2 \left[
\frac{1}{U^2 \sin^2(K(x+y)) + (2 k D_0)^2 } \right. \right.
\nonumber \\
& & \left. \left.
+ \frac{1}{U^2 \sin^2(K(x-y)) + (2 k D_0)^2 } \right] \right\}
+ O(\varepsilon^3)
\label{eq:3.29}
\end{eqnarray}

In Fig.~\ref{fig:1} we compare the exact multiple-scale solution 
 for $D^E_{xx}(y)$ in the case of the large-scale shear flow,
with the approximation~(\ref{eq:3.28}) and the constant estimation 
based on the sole small-scale cellular flow, respectively. 
In most of 
the domain the first-order approximation recovers quite well 
the exact solution, with the exception of narrow regions where 
the large-scale flow vanishes and the actual diffusivity is mainly 
determined by the cellular flow. 

In all figures and tables we show quantities made dimensionless 
in the form: 
\begin{equation}
x \to \frac{x}{L_0}\;, v \to \frac{v}{U_0}\;, t \to t \frac{U_0}{L_0}\;, D \to \frac {D}{L_0 U_0}
\label{eq:adim}
\end{equation}
where $U_0 = U$ and $L_0 = L/2\pi$.

\begin{figure}[pt]
\begin{center}
\includegraphics[scale=0.7]{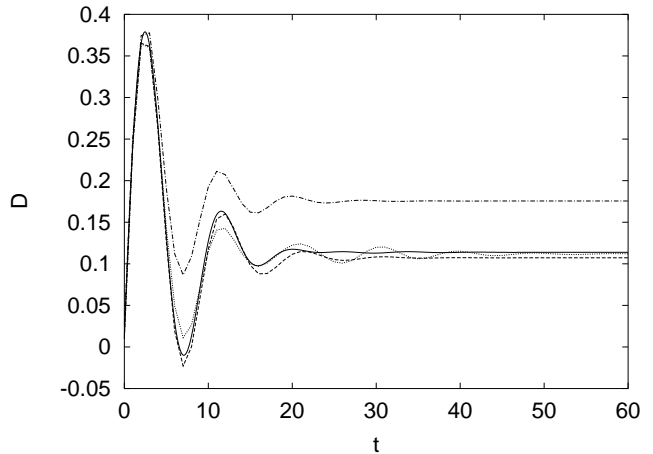}
\caption{The same as in Fig.~\ref{fig:2} for the case of the 
large-scale cellular flow ($\ell / L  = 1/4$,$u / U = 1/4$, $D_0 = 10^{-2}$).
Time evolution of the asymptotic eddy-diffusivity
$D^{\mathcal L}$  is well approximated by the 
first-order approximation~(\ref{eq:3.6}) (solid line), 
while the ``naive estimation'' (dash-dotted) does not match
the actual values which depend on the 
relative phase-shifts between the two fields:
case $a$ dashed line; case $b$ dotted line.
Units are made dimensionless according to Eq.~(\ref{eq:adim}).
}
\label{fig:3}
\end{center}
\end{figure}

Once the first-order approximations~(\ref{eq:3.28}--\ref{eq:3.29}) for the 
eddy-diffusivity have been plugged into the pre-asymptotic
large-scale equation, we compute the asymptotic eddy-diffusivity
at very large scales $\mathcal{L}$.\\ 
Numerical integration of the auxiliary equation~(\ref{eq:3.13})
is advanced in time until the asymptotic eddy-diffusivity
given  by Eq.~(\ref{eq:3.12}) converges to its constant value. The latter 
is then compared with the values given by
homogenization of equation~(\ref{FP}) for different phase-shifts between
${\bm U}$ and ${\bm u}$.\\
The observed variability of  $D^{{\mathcal L},ex}$ for different phase-shifts
provides an estimation of the accuracy of the multiple-scale results.
Indeed, as already noted, the two successive homogenizations
do  not capture any  effect induced by the phase-shift. \\
Here, we consider the two extreme possibilities: i) 
the zeros of the large-scale flow coincide  with the nodes of the 
small-scale cellular flow (case a); ii) the zeros 
of the large-scale flow  are located on the bellies of the small-scale 
cellular flow (case b).\\
In addition, we compute the constant eddy-diffusivity 
$D^E_{ij} = \tilde{D} \delta_{i j}$ of the velocity field
containing the small-scale cellular flow only. 
This leads to a rather crude approximation 
for the asymptotic eddy-diffusivity $D^{{\mathcal L},n}$
(in the following, we will  refer to as ``naive approximation'').\\
For the large-scale shear, the asymptotic diffusion tensor $D^{\mathcal L}$
is diagonal and strongly anisotropic.
In Fig.~\ref{fig:2} we show its component
$D^{\mathcal L}_{xx}$ in the direction parallel to the large-scale shear flow. 
The scale separation is $\ell/L  = 1/4$, 
the ratio of amplitudes is $u/U = 1/4$ and 
the molecular diffusion is fixed to the value $D_0 = 10^{-2}$.

In the direction of the shear the effect of the small-scale flow is to 
reduce the asymptotic diffusion coefficient, which for the  
pure large-scale shear flow would be given by 
\begin{equation}
D^{\mathcal L}_{xx} = D_0 + \frac{1}{2} \frac{U^2}{K^2 D_0} = 50.01 \;.
\end{equation}
Such reduction is due to interference mechanisms between small-scale and 
large-scale motion \cite{MV97}.\\
With our parameters the actual reduction is of the order of $ 20 - 30\% $, 
depending on the phase-shift between $\bm{U}$ and $\bm{u}$.
The first-order approximation (\ref{eq:3.6})
for the eddy-diffusivity provides 
a good estimation giving a reduction for $D^{\mathcal L}_{xx}$ of about
$28 \% $ (see Tab.~\ref{tab1}).
On the contrary the ``naive approximation'' gives a reduction for 
$D^{\mathcal L}_{xx}$ of about $60\% $,
which is deeply wrong. 

In the transverse direction, the bare molecular diffusivity $D_0$ 
is increased by the presence of the small-scale flow. 
The ``naive approximation'' overestimates this 
effect, giving an enhancement  of about $170 \%$ of $D_0$, while 
approximation~(\ref{eq:3.6}) is in rather good agreement with 
the actual value of about $12 - 20 \%$. 

It is worth stressing that the large errors given by the 
``naive approximation'' rather than being consequences of finite 
scale separation are mainly due to the fact that 
the effects of the large-scale flow has been neglected
in the constant eddy-diffusivity $D^{E}_{i j}({\bm X},T) = 
\tilde{D} \delta_{i j}$. Indeed,
with a larger scale separation $\ell/L = 1/8$ 
the approximate solution gives results within the $2 \%$ 
of the actual values, while the ``naive approximation'' 
still gives an error of about $30 \%$ (see Tab.~\ref{tab1}). 

In the case of the large-scale cellular flow (see Fig.~\ref{fig:3})
the asymptotic eddy-diffusivity is isotropic, and 
the first-order approximation is even more robust, providing 
good estimations also for $\epsilon = \ell / L = 1/4$ 
and $\varepsilon = u / U = 1/2$ 
(see Tab.~\ref{tab2}). 
The errors of the ``naive approximation'' are of the 
order of $100 \%$.

\begin{table}[t]
\begin{center}
\begin{tabular}{|l|l|l|l|l|} 
\hline
$\ell /L$ & 
$u/U$ & 
$D^{{\mathcal L},ex}$ & 
$D^{\mathcal L}$ & 
$D^{{\mathcal L},n}$ \\
\hline
$1/4$ & 
$1/4$ & 
$D_{xx} = 41.6^{(a)} $-$ 34.5^{(b)}$ &
$36.7$ &
$18.7$ \\
& & 
$D_{yy} = 0.0112^{(a)} $-$ 0.0122^{(b)}$ & 
$0.0119$ &
$0.0267$ \\
\hline
$1/8$ & 
$1/4$ & 
$D_{xx} = 41.5^{(a)} $-$ 40.5^{(b)}$ &
$39.6$ &
$28.3$ \\
& & 
$D_{yy} = 0.0112^{(a)} $-$ 0.0113^{(b)}$ & 
$0.0115$ &
$0.0178$ \\
\hline
\end{tabular}
\caption{Asymptotic eddy-diffusivity resulting from the effects of 
large-scale shear flow ($U=1$, $L= 2 \pi$), small-scale cellular flow
and molecular diffusivity $D_0=0.01$. 
$D^{{\mathcal L},ex}$ (cases $(a)$ and $(b)$) are the actual values
obtained from direct homogenization 
of the whole velocity field ${\bm v} = {\bm U} +{\bm u}$.
$D^{\mathcal L}$, 
and $D^{{\mathcal L},n}$ are obtained from the homogenization of 
the pre-asymptotic equation where the pre-asymptotic 
eddy-diffusivities are approximated by expression~(\ref{eq:3.6})
and by retaining the sole small-scale cellular flow, respectively.
Units are made dimensionless according to Eq.~(\ref{eq:adim}).
}
\label{tab1}
\end{center}
\end{table}
\begin{table}[t]
\begin{center}
\begin{tabular}{|c|c|c|c|c|} 
\hline
$\ell /L$ & 
$u/U$ & 
$D^{{\mathcal L},ex}$ &  
$D^{\mathcal L}$ &  
$D^{{\mathcal L},n}$ \\
\hline
$1/4$ & 
$1/2$ & 
$0.111^{(a)}$ - $0.123^{(b)}$ &
$0.135$ &
$0.209$ \\
\hline
$1/4$ & 
$1/4$ & 
$0.107^{(a)}$ - $0.112^{(b)}$ &
$0.113$ &
$0.175$ \\
\hline
\end{tabular}
\caption{The same as in Tab.~\ref{tab1} for the large-scale cellular flow.}
\label{tab2}
\end{center}
\end{table}

\subsection{An empirical ``recipe''}

We discuss now an empirical
``recipe'' to obtain a constant (i.e. having no variation in space and in time)
eddy-diffusivity to describe pre-asymptotic scales.
The question is thus 
on whether it is possible to mimic the pre-asymptotic  
transport by means of
an average diffusion tensor $D^{E,a}_{ij}$ 
which still takes into account 
the effects of the large-scale flow, ${\bm U}$, but does not depend 
on the position. In general it is not clear which is the correct way for  
averaging $D_{ij}({\bm X}, T)$ to obtain a constant, but 
still anisotropic diffusion tensor.  
Here, we propose a possible ``recipe'' which is inspired 
by the multiple scale approach. 
The idea consists in applying the homogenization technique 
just on the diffusive term of the pre-asymptotic equation, 
obtaining $D^{E,a}_{ij}$ in the same way of  
$D^{\mathcal L}$:
\begin{equation} 
D^{E,a}_{ij} = \frac{\langle D_{ik} \partial_k \chi_j \rangle +
   \langle D_{jk} \partial_k \chi_i \rangle}{2}
+ \frac{\langle D_{ij} \rangle + \langle D_{ji} \rangle}{2} \;,
\label{tempt1}
\end{equation}
where the vector field ${\bm \chi}$ is solution
of the auxiliary equation
\begin{equation}
\partial_t \chi_k + ({\bm U} \cdot {\bm \partial} ) \chi_k 
- \partial (D_{ij} \partial_j \chi_k) =  \partial_i D_{ik} \;.
\label{tempt2}
\end{equation}
Although the recipe~(\ref{tempt1}-\ref{tempt2}) cannot 
be rigorously proved, it is possible to give
a rough argument in favor of it.
Eqs.~(\ref{tempt1}-\ref{tempt2}) can be seen as the analogous of 
Eqs.~(\ref{eq:3.12}-\ref{eq:3.13}) in which only the  
eddy-diffusivity contributions to the asymptotic diffusion tensor 
have been retained.\\
The above discussed pre-averaged constant diffusion tensor 
is potentially interesting in applications, where it 
is almost impossible to deal with space-dependent eddy-diffusivities. 
Let us stress the fact that $D^{E,a}_{ij}$ in~(\ref{tempt1}) 
is constant, but it takes into account the effects 
of the large-scale flow to provide a correct estimation 
of an effective diffusion tensor.\\ 
Numerical simulations of the pre-asymptotic equation~(\ref{lsse})
in which $D_{ij}({\bm X}, T)$ is replaced by the constant 
tensor $D^{E,a}_{ij}$ confirms that 
this averaging recipe leads to considerable improvements 
with respect to the ``naive approximation'' 
obtained without considering the effects of the large-scale flow.

Table~\ref{tab3} shows the results 
in the case of the large-scale shear flow, where this averaging leads 
to a rather good approximation $D^{{\mathcal L},a}$
for the asymptotic eddy-diffusivity; 
similar results holds for the case of large-scale cellular flow.

\begin{table}[t]
\begin{center}
\begin{tabular}{|l|l|l|l|l|} 
\hline
$\ell /L$ & 
$u/U$ & 
$D^{{\mathcal L},ex}$ & 
$D^{{\mathcal L},a}$ & 
$D^{{\mathcal L},n}$ \\
\hline
$1/4$ & 
$1/4$ & 
$D_{xx} = 41.6^{(a)} $-$ 34.5^{(b)}$ &
$42.1$ &
$18.7$ \\
& & 
$D_{yy} = 0.0112^{(a)} $-$ 0.0122^{(b)}$ & 
$0.0118$ &
$0.0267$ \\
\hline
$1/8$ & 
$1/4$ & 
$D_{xx} = 41.5^{(a)} $-$ 40.5^{(b)}$ &
$43.6$ &
$28.3$ \\
& & 
$D_{yy} = 0.0112^{(a)} $-$ 0.0113^{(b)}$ & 
$0.0115$ &
$0.0178$ \\
\hline
\end{tabular}
\caption{The same as in Tab.~\ref{tab1}. 
The asymptotic eddy-diffusivity 
$D^{{\mathcal L},a}$ is obtained from the homogenization of 
the pre-asymptotic equation where the pre-asymptotic 
eddy-diffusivity is approximated by the constant value given 
by Eq.~(\ref{tempt1}).  
}
\label{tab3}
\end{center}
\end{table}

\section{Multiple-scale expansion and Renormalization Group}
\label{rg}
In previous sections we studied the problem of large-scale transport
in field varying on two separated scales that we called, large and
small scales, respectively.\\
In practical applications, one has to deal with advecting velocity
fields having almost a continuous of active scales.
In this latter cases, we can write
\begin{equation}
  \bm{u}(\bm{x},t) =\sum_{n=0}^N
\bm{u}_n(\bm{x},t)= \bm{u}_0(\bm{x},t)+\delta \bm{u}(\bm{x},t)
\end{equation}
where the Fourier transform of $\bm{u}_n(\bm{x},t)$ is picked on
 wave-numbers around $k_n \sim l_n^{-1}=2^{-n}l_0^{-1}$.\\
Denoting with  $E(k)$  the energy spectrum, one has
\begin{equation} \frac{1}{2} \langle |\bm{u}_n(\bm{x},t)|^2 \rangle \simeq
 \int_{k_n}^{k_{n+1}} E(k) dk \,\,\, .
\end{equation}

We are now ready to address the following question: what is the effect of
$\delta \bm{ u}(\bm{ x},t)$ on the effective, asymptotic eddy-diffusivity?
 In other words, we aim at obtaining an effective large-scale equation
and  determine the dependence of $\bm{U}^{E}$
and $D^{E}$ on $\delta \bm{ u}(\bm{ x},t)$ and $D_0$, respectively.\\
A natural way to answer our question is to exploit the 
renormalization group point of view. The basic idea proceeds along
this steps:
\begin{enumerate}
\item  starting from the original equation (\ref{FP}):
one considers the field
\begin{equation}
\bm{U}_{N-1}(\bm{ x},t) =\sum_{n=0}^{N-1}\bm{ u}_n(\bm{ x},t)
\end{equation}
as the the one at large scales and $\bm{ u}_N(\bm{ x},t)$
as the contribution at small scales.
Recalling 
the results of the multiple-scale expansion reported in Sec.~\ref{mse},
we can write the effective equation for the
field including the contribution up to the scale $N-1$, i.e.
\begin{equation}
 \partial_t \theta + {\bm U}^E_{N-1}\cdot
\bm{\nabla}\theta =
\nabla (D^E_{N-1} \nabla  \theta)  \,\, ,
\end{equation}
where $\bm{U}^E_{N-1}$ and  $D^E_{N-1}$ are determined by the
multiple-scale analysis of Secs.~\ref{mse} and~\ref{pertu}.\\
It is rather obvious that it is almost impossible to repeat
in full details the multiple-scale procedure. On the other hand,
if one is interested to the sole order of magnitudes, interesting
results can be obtained by neglecting the dependence on $\bm{ x}$.
In this spirit we obtain:
\begin{equation}
D^E_{N-1} \simeq D_0+ const.
\frac{ {D_0 \langle |{\bm u}_N|^2 \rangle k_N^2}}{ {(D_0 k_N^2)^2 +(k_N |{\bm U}_{N-1}|)^2}}
\end{equation}
and
\begin{equation}
{\bm U}^E_{N-1}= {\bm U}_{N-1}({\bm x},t) +
\delta {\bm U}_{N-1} 
\end{equation}
where $\delta {\bm U}_{N-1}$ is the compressible
contribution originated from the dependence of  $D^E_{N-1}$ on ${\bm x}$.

\item
As a second step, one now has to iterate the previous procedure.
In order  to simplify the computation, as before we
do not take into account neither the dependence of $D^E_{N-1}$ on $\bm{x}$
nor the compressible correction on ${\bm U}^E_{N-1}$.
We have just to replace $D_0$ with $D^E_{N-1}$, 
${\bm U}_{N-1}$ with ${\bm U}_{N-2}$, $k_N$ with $k_{N-1}$ and so on.
When doing so, we arrive at
\begin{equation}
D^E_{N-2} \simeq D^E_{N-1} + const.
\frac{ {D^E_{N-1}
\langle |{\bm u}_{N-1}|^2 \rangle k_{N-1}^2}}{ 
{(D^E_{N-1} k_{N-1}^2)^2 +(k_{N-1} |{\bm U}_{N-1}|)^2}}
\label{recursion}
\end{equation}
and similarly for ${\bm U}^E_{N-2}$, and so on
for $N-3, N-4, ...$.\\
\end{enumerate}
The effective asymptotic eddy-diffusivity $D^E$ is obtained by iterating
the recursive relation (\ref{recursion}).
Two interesting limits have been identified:

i) the dominant term in the denominator of (\ref{recursion}) is
$(D^E_{N-1} k_{N-1}^2)^2$ and the recursive formula becomes
\begin{equation}
D^{E}_{N-2} \simeq D^{E}_{N-1} + const.
\frac{ \langle |{\bm u}_{N-1}|^2 \rangle }{D^{E}_{N-1} k_{N-1}^2}  \, ;
\label{rel1}
\end{equation}

ii) the dominant term in the denominator of (\ref{recursion}) is
$(k_{N-1} |{\bm U}_{N-1}|)^2$ and we thus have
\begin{equation}
D^{E}_{N-2} \simeq 
D^{E}_{N-1} \left( 1 + 
const. \frac{ \langle |{\bm u}_{N-1}|^2 \rangle }{|{\bm U}_{N-1}|^2}
\right).
\label{rel2}
\end{equation}
The relation (\ref{rel1})  
coincides with the result obtained by Moffatt~\cite{12}. 
Iterating  (\ref{rel1}) one easily obtain:
\begin{equation}
  D^E \sim {\sqrt{ \int k^{-2} E(k) d k}}
\end{equation}
i.e. an eddy-diffusivity which does not depend on the molecular diffusivity 
$D_0$.\\
On the contrary, exploiting the fact that $\bm{U}_{N_1} \simeq \bm{u}_0$
from (\ref{rel2})  one has:
\begin{equation}
D^E \sim D_0 
\left(1+ const. \sum_n \frac{\langle |{\bm u}_n|^2 \rangle}{|\bm{u}_0|^2}
\right) \propto D_0 
\end{equation}
In summary, from the iteration of 
the recursive rule~(\ref{recursion})
one can obtain at least two fixed points. In the first case the 
asymptotic eddy-diffusivity is determined only from the velocity 
field and it does not depend on $D_0$. This allows for values of $D^E$
much larger than $D_0$. In the second limit, one has a small variation of 
the asymptotic eddy-diffusivity which remains of the same order  of $D_0$. 
  
\section{Conclusions}
\label{conclu}
We have investigated both analytically and numerically the
pre-asymptotic transport of a passive scalar field
on large scales, say, of order $L$.
The velocity field advecting the scalar is formed by
a large-scale component ${\bm U}$ varying on
scales of order of $L$ and by a small-scale fluctuation, ${\bm u}$,
which varies  on scale of order of $\ell$ much smaller than $L$. 
The presence of a small
parameter $\ell/L$ naturally allows a perturbative analysis: the
so-called multiple-scale strategy.\\
The following results must be emphasized.
\begin{enumerate}

\item Pre-asymptotic scalar transport is ruled by a Fokker-Planck
equation involving an effective eddy-diffusivity field and an
effective advecting velocity. Although explicit expressions for 
such effective fields cannot  be determined in general, nevertheless
it is apparent that the eddy-diffusivity does depend on the
large-scale advecting velocity. This is in contrast with the usual
point of view which sees the eddy-diffusivity  as the cumulative 
result  of interactions involving the sole small scales.
Such aspect can be rather relevant in a geophysical context~\cite{future}

\item If one does the additional assumption that small-scale
fluctuations are sufficiently weaker than the large-scale fluctuations
(i.e. $u/U \ll 1$), an approximate explicit expression for the
eddy-diffusivity tensorial field can be obtained. From such expression
it becomes explicit  the dependence of the eddy-diffusivity
on the large-scale velocity, which, in turns, carries a
spatio-temporal dependence on large scales.\\ 

\item If the small-scale velocity 
$\bm{u}$ has defined parity under spatial/temporal inversion,
the sole symmetric part of $D_{i j}$ is relevant for the pre-asymptotic
dynamics. The same conclusion holds if $\bm{u}$ is a small-scale
stationary, homogeneous and isotropic turbulent field. 

\item We have tested numerically the validity of our approximated
expression for the eddy-diffusivity for value of $u/U  $ and $\ell/L$
not necessarily much less the unity. As expected, the range of 
reliability of our approximation extends to finite value of the
above ratios. This seems an important conclusion for applications
in the realm of geophysics and oceanography.

\item Exploiting the explicit formula for the eddy-diffusivity,
we have presented a generalization of our results to situations
with a continuous of active scales. This procedures gives rise to a
sort of renormalization group through which it is possible to extract 
two completely different regimes of transport. 
\end{enumerate}

We would like to conclude with a short discussion on the applicability
of our results and, more generally, of
multiple-scales techniques to geophysical problems. 
As far as the first point is concerned, a paradigmatic example of
a possible application
is provided by the investigation of pollutant dispersion in 
Planetary Boundary Layer (PBL). The latter is  
a thin ($\sim 1000\ m$) atmospheric
 layer near the ground, where the airflow is strongly driven by
 sink/source forcing terms arising from the bottom boundary, e.g.
 due to the orography.  The decomposition of the velocity field as
 ${\bm v} = {\bm u} + {\bm U}$, ${\bm u}$ being a fluctuating random
 component, whose statistical properties are prescribed, and
 ${\bm U}$ is a slowly-varying part
 is a standard decomposition. In way of example, 
the slow component ${\bm U}$ describes synoptic variations while
the fast component ${\bm u}$ modelizes, for instance, orographic
excitations.\\
Let us now point out some important limitations in the applicability of 
the multiple-scale analysis to geophysical problems.
A first obvious  limit comes from the separation 
between the characteristic scales of the flow. 
The multiple-scale approach is  
strictly valid only in the case of large separation, 
while the typical separation of scales and amplitudes 
in realistic geophysical flows is not very large. 
Actually, this does not seem a severe restriction, 
since the results obtained in the limit of infinite separation 
provide rather good approximations also valid for moderate separations
(see e.g. the numerical results of Sec.~\ref{num}).\\
Moreover, the multiple-scale approach requires
a detailed knowledge of the Eulerian velocity field, 
which is not always available experimentally.
It thus seems to us that an attempt to build a pre-asymptotic 
equation for the transport, using only Lagrangian experimental data,  
should be a further important step toward a satisfactory understanding on
how to modelize large-scale transport in geophysical flows. 

\section*{Acknowledgments}
This work has been supported by Cofin 2003
``Sistemi Complessi e Problemi a Molti Corpi''.
Numerical simulations have been performed at CINECA (INFM parallel
computing initiative).
                                                                               

\end{document}